\documentclass[aps,prb,showpacs,reprint,groupedaddress]{revtex4-1}

\usepackage[final]{graphicx}
\usepackage{amsmath,amssymb,latexsym,bm,here}
\usepackage{wasysym}
\usepackage{amsfonts}
\usepackage{color}
\usepackage{nicefrac}
\usepackage[separate-uncertainty]{siunitx}
\usepackage{booktabs,dcolumn,tabularx}
\usepackage[bookmarks=false,pdffitwindow=true,colorlinks,linkcolor=black,citecolor=black,filecolor=black,urlcolor=blue,pdfborder={0 0 0}]{hyperref}
\usepackage[english]{babel}

\def\Tensor#1{\overset{\text{\tiny$\bm\leftrightarrow$}}{#1}}

\renewcommand{\vec}[1]{\mathbf{#1}}

\begin{document}

\preprint{APS/123-QED}

\title{Second-harmonic phonon spectroscopy of $\alpha$-quartz}

\author{Christopher J. Winta}

\author{Sandy Gewinner}

\author{Wieland Sch\"ollkopf}

\author{Martin Wolf}

\author{Alexander Paarmann}
\email{alexander.paarmann@fhi-berlin.mpg.de}

\affiliation{Fritz-Haber-Institut der Max-Planck-Gesellschaft, Department of Physical Chemistry, Faradayweg 4-6,
14195 Berlin, Germany}

\date{\today}

\begin{abstract}
We demonstrate midinfrared second-harmonic generation as a highly sensitive phonon spectroscopy technique that we exemplify using $\alpha$-quartz (SiO$_2$) as a model system. A midinfrared free-electron laser provides direct access to optical phonon resonances ranging from \SIrange{350}{1400}{\per\centi\meter}. While the extremely wide tunability and high peak fields of a free-electron laser promote nonlinear spectroscopic studies---complemented by simultaneous linear reflectivity measurements---azimuthal scans reveal crystallographic symmetry information of the sample. Additionally, temperature-dependent measurements show how damping rates increase, phonon modes shift spectrally and in certain cases disappear completely when approaching $T_c=\SI{846}{\kelvin}$ where quartz undergoes a structural phase transition from trigonal $\alpha$-quartz to hexagonal $\beta$-quartz, demonstrating the technique's potential for studies of phase transitions.
\end{abstract}


\maketitle


\section{Introduction}

Nonlinear optical spectroscopy has proven to be a powerful tool to study crystalline solids,\cite{Shen2003} offering additional experimental degrees of freedom compared to linear optical techniques since the symmetry properties of the crystal enter into the nonlinear signals through the nonlinear susceptibility tensor,\cite{Li2013,Zhao2015,Fiebig2005} providing unique insight into both, crystallographic and electronic symmetries of the given system.\cite{Zhao2015,Bovino2013,Fiebig2005,Niedermeier1999,Ohno2016} Additionally, nonlinear approaches often provide improved contrast and sensitivity,\cite{Paarmann2016a} in particular upon symmetry changes across order-to-order phase transitions. While employed extensively for the study of electronic transitions,\cite{Zhao2015,Fiebig2005} this idea could be similarly useful when considering the symmetry properties of zone-center optical phonons in polar crystals. Here, linear optical techniques like reflectance spectroscopy typically only provide access to frequency, amplitude and linewidth changes of the phonon resonance.\cite{Lobo2007,Schleck2010}\par
For nonlinear optical studies of optical phonons, vibrational sum-frequency generation (SFG) spectroscopy is the most well-established technique\cite{Shen1989}, which, due to its even-order, can probe non-centrosymmetric media\cite{Liu2008} or surface- and interface phonons in inversion-symmetric crystals.\cite{Shen1994,Bonn2000,Liu2008,Tong2015a} In principle, second-harmonic generation (SHG) constitutes an attractive alternative even-order technique which is widely used in the visible and near-infrared.\cite{Shen1994,Yamada1994,Niedermeier1999,Bovino2013,Fiebig2005,Li2013a,Zhao2015,Ohno2016} However, only few studies were performed investigating optical phonon resonances in the midinfrared (mid-IR) to terahertz (THz) spectral region,\cite{Mayer1986,Dekorsy2003,Paarmann2015,Paarmann2016a} owing to the scarcity of respective intense and narrowband laser sources and the lack of single-photon detectors in the infrared. Mid-IR free-electron lasers (FELs) are in fact highly suitable for such investigations thanks to their broad wavelength tunability, narrow bandwidth and high peak power.\cite{Paarmann2015,Schollkopf2015}\par
Notably, for the study of optical phonon resonances, SHG in the mid-IR has several potential advantages over SFG: (i) higher symmetry of the nonlinear susceptibility tensor,\cite{Shen2003} (ii) improved phonon enhancement due to doubly resonant mid-IR excitation,\cite{Paarmann2016a} and (iii) access to higher order resonances and different mode symmetries due to different selection rules.\cite{Flytzanis1972} To explore these mechanisms and evaluate the potential of mid-IR SHG as a phonon spectroscopy, we have chosen a suitable model system that is well-studied with SFG and other vibrational spectroscopy techniques.

The lattice dynamics of quartz has been subject to various studies which include SFG,\cite{Liu2008} Raman\cite{Krishnan1945,Ichikawa2003} and IR spectroscopy\cite{Spitzer1961} as well as neutron\cite{Dorner1980,Strauch1993} and X-ray scattering.\cite{Bauer1971} 
Mostly due to its piezoelectricity, quartz is of great technological importance,\cite{Bosak2012} while at the same time constituting a well-studied model system for nonlinear optical techniques.\cite{Liu2008,Ohno2016} Its broken inversion symmetry supports even-order nonlinear processes, in particular formidable SHG yields, and its numerous vibrational modes present a rich test ground for phonon spectroscopies.

Here, we introduce mid-IR SHG phonon spectroscopy and exemplify the technique for $\alpha$-quartz (SiO$_2$). Making use of the large tunability of the FEL, we are able to investigate essentially all optical phonon resonances of $\alpha$-quartz,\cite{Strauch1993} spanning a broad frequency range from \SIrange{350}{1400}{\per\centi\meter}. We observe an enhancement of the SHG yield over several orders of magnitude at numerous phonon resonances, well correlated with characteristic features in the reflectivity spectrum which is measured simultaneously. Additionally, the trigonal crystal structure of the sample causes a pronounced azimuthal angle-dependence of the SHG signal which we use to gain information about the contributing $\chi^{(2)}$ tensor elements. Finally, temperature-dependent measurements indicate that second-harmonic phonon spectroscopy is highly sensitive to the $\alpha\rightarrow\beta$ phase transition of quartz.


\section{Experiment}

\begin{figure}[htb]
\includegraphics[width=\linewidth]{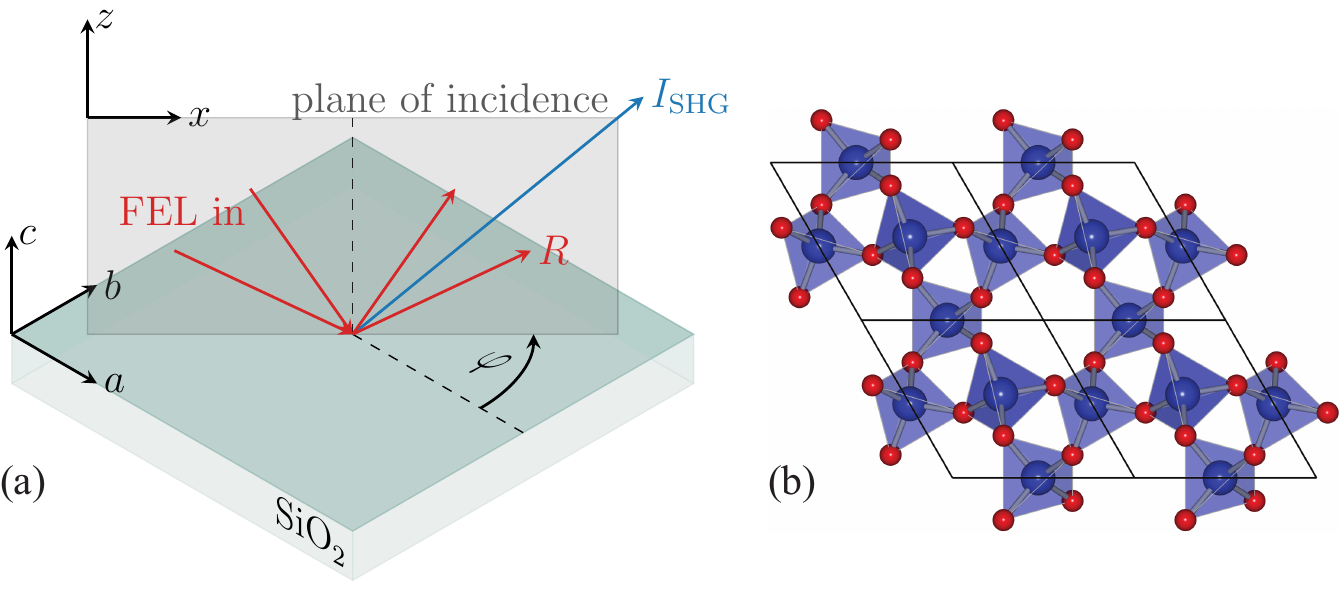}
\caption{(a): Schematic of the experimental setup and definition of the coordinate systems. Non-collinear two-beam excitation with the FEL generates two-pulse correlated SHG in reflection. Rotation of the sample about the $z$-axis provides the azimuthal behavior of the SHG signal. (b): Crystal structure of $\alpha$-quartz (view along the optic $c$-axis). Oxygen atoms (red) form tetrahedra around the silicon atoms (blue).}
\label{fig:setup}
\end{figure}

The experimental arrangement and concept of mid-IR SHG are described in detail elsewhere.\cite{Paarmann2015,Paarmann2016a} In short, our experimental setup (Fig.~\ref{fig:setup}) resembles a non-collinear autocorrelator geometry in reflection where two focused FEL beams impinge on the $\alpha$-quartz sample at incidence angles of \ang{30} and \ang{60}, respectively. They generate the spatially and temporally correlated SHG signal which emerges between both reflected excitation beams and is detected by a liquid nitrogen cooled mercury cadmium telluride/indium antimonide (MCT/InSb) sandwich detector (InfraRed Associates). Additionally, the sample is mounted on a motorized rotation stage (Newport) which allows to turn the sample about the surface normal and thereby facilitates scans of the azimuthal angle $\varphi$. In order to prevent light at the fundamental wavelength to be scattered onto the detector, a variety of short-pass spectral filters is used, comprising MgF$_2$, CaF$_2$ (Thorlabs) and ZnS/ms (Korth) windows, depending on the spectral range investigated. Additionally, \SIlist{7;9;13.5}{\micro\meter} long-pass filters (LOT) are used selectively to block intrinsic higher harmonics generated by the FEL itself.\par

Details on the FEL are given elsewhere.\cite{Schollkopf2015} In short, the electron gun is operated at a micropulse repetition rate of \SI{1}{\giga\hertz} with an electron macropulse duration of \SI{10}{\micro\second} and a macropulse repetition rate of \SI{10}{\hertz}. For these measurements, the electron energy was set to \SI{23.5}{\mega\electronvolt} or \SI{31}{\mega\electronvolt}, allowing to tune the FEL output wavelength $\lambda$ from $\sim$\SIrange{12}{28}{\micro\meter} (\SIrange{350}{850}{\per\centi\meter}) and from $\sim$\SIrange{7}{18}{\micro\meter} (\SIrange{550}{1400}{\per\centi\meter}), respectively, by varying the motorized undulator gap. In order to display the resulting measurements in one spectrum covering the whole frequency range from \SIrange{360}{1400}{\per\centi\meter}, multiple overlapping measurements with different FEL electron energies and optical filter configurations were merged. The cavity desynchronism $\Delta L$ is set to $2\lambda$, causing narrow-band operation\cite{Schollkopf2015} with typical full-width-at-half-maximum (FWHM) of $\sim$0.5\%. Polarization rotation by \ang{90} of the linearly P-polarized FEL beam is achieved by two subsequent wire-grid polarizers (Thorlabs) set to \ang{45} and \ang{90} with respect to the incoming polarization, respectively. A third polarizer in front of the detector allows to selectively measure a specific polarization component of the SHG signal. Scanning the FEL frequency $\omega=2\pi c/\lambda$ via the FEL undulator gap results in a spectroscopic measurement of the SHG intensity, while scanning the angle $\varphi$ at fixed $\omega$ provides the azimuthal SHG behavior. Additionally, the reflected fundamental beam incident at \ang{60} is detected with a pyroelectric detector, allowing to simultaneously obtain the IR reflectivity spectra.\par
The sample investigated here is an optically polished single crystal $c$-cut window of $\alpha$-quartz (Crystal GmbH) with the [0001] crystal axis, i.e. the optic $c$-axis, perpendicular to the surface. For temperature-dependent measurements, the sample was mounted on a sample heating stage allowing for a restricted azimuthal rotation ($\sim\ang{100}$) for temperatures up to \SI{1025}{\kelvin}. We note that the temperature sensor was placed on the backside of the sample, leading to an overestimation of the actual sample temperature.


\section{Theoretical Description}
\label{sec:theory}
The general theory of SHG is well-established.\cite{Shen1989} Here, we specifically treat resonant mid-IR SHG in reflection with non-collinear, correlated two-pulse excitation (see Fig.~\ref{fig:setup}). Two tunable IR beams, both at frequency $\omega$, with incoming wave vectors $\vec{k}_1^\mathrm{i}$ and $\vec{k}_2^\mathrm{i}$ at angles of incidence $\alpha_1^\mathrm{i}=\ang{30}$ and $\alpha_2^\mathrm{i}=\ang{60}$, respectively, impinge on the sample. The generated second-harmonic radiation at frequency $2\omega$ emerges at an angle \mbox{$\alpha_\mathrm{SHG}^\mathrm{r}=\arcsin\left[\left(\sin\alpha_1^\mathrm{i}+\sin\alpha_2^\mathrm{i}\right)/2\right]\approx\left(\alpha_1^\mathrm{i}+\alpha_2^\mathrm{i}\right)/2$}. For crystals with broken inversion symmetry, the surface contribution to the second-order nonlinear signal is typically negligible.\cite{Shen1989} Therefore, the symmetry properties of the second-order nonlinear susceptibility tensor $\Tensor{\chi}^{(2)}$ are given by the bulk crystal symmetry and we here solely consider the bulk SHG polarization of the form
\begin{align}
\begin{split}
\vec{P}(2\omega)\propto \Tensor{\chi}^{(2)}&(2\omega,\omega,\omega):\\ &\left(\Tensor{L}_1(\omega)\vec{E}_1(\omega)\right)\left(\Tensor{L}_2(\omega)\vec{E}_2(\omega)\right),
\label{eq:sh_polarization}
\end{split}
\end{align}
where $\Tensor{L}_{1(2)}(\omega)$ is the Fresnel transmission tensor\cite{Paarmann2015,Mosteller1968} for the first (second) incident beam, accounting for macroscopic local field corrections,\cite{Shen1989} and $\vec{E}_{1(2)}(\omega)$ its respective incident electric field vector. Projecting the nonlinear polarization onto the electric field direction of the reflected SHG beam, $\hat{e}_\mathrm{SHG}$, and considering the Fresnel transmission of the nonlinear polarization components  at $2\omega$ back into air gives the reflected SHG intensity,\cite{Shen1989}
\begin{equation}
I(2\omega)\propto\left|\left(\Tensor{\widetilde{L}}_\mathrm{SHG}(2\omega)\vec{P}(2\omega)\right)\cdot\hat{e}_\mathrm{SHG}\right|^2/\Delta k^2,
\label{eq:sh_intensity}
\end{equation}
where $\Tensor{\widetilde{L}}_\mathrm{SHG}$ denotes the Fresnel tensor for the reflected SHG beam coming out of the sample into air and $\hat{e}_\mathrm{SHG}$ its field direction. Additionally, \mbox{$\Delta k^2=\left|\vec{k}^\mathrm{SiO{_2}}_\mathrm{SHG}-\vec{k}^\mathrm{SiO{_2}}_1-\vec{k}^\mathrm{SiO{_2}}_2\right|^2$} accounts for the wave vector mismatch in reflection, with $\vec{k}^\mathrm{SiO{_2}}_{1(2)}$ being the first (second) transmitted fundamental and $\vec{k}^\mathrm{SiO{_2}}_\mathrm{SHG}$ the reflected SHG wave vector inside the crystal, respectively.\par
The Fresnel transmission tensor $\Tensor{L}(\omega)$ which determines the local fields inside the sample, is diagonal and its elements are highly dispersive.\cite{Paarmann2015} These are straightforwardly derived from Maxwell's equations and for uniaxial crystals with the optic axis along $z$ read:

\begin{align}
\begin{split}
L_{xx}(\omega,\alpha^\mathrm{i}) &= \frac{2k_z^\mathrm{SiO{_2},e}(\omega,\alpha^\mathrm{i})}{\varepsilon_\perp(\omega)k_z^\mathrm{air}(\omega,\alpha^\mathrm{i})+k_z^\mathrm{SiO{_2},e}(\omega,\alpha^\mathrm{i})},\\
L_{yy}(\omega,\alpha^\mathrm{i}) &= \frac{2k_z^\mathrm{air}(\omega,\alpha_\mathrm{i})}{k_z^\mathrm{air}(\omega,\alpha^\mathrm{i})+k_z^\mathrm{SiO{_2},o}(\omega,\alpha^\mathrm{i})},\\
L_{zz}(\omega,\alpha^\mathrm{i}) &= \frac{\varepsilon_\perp(\omega)}{\varepsilon_\parallel(\omega)}\ \frac{2k_z^\mathrm{air}(\omega,\alpha^\mathrm{i})}{\varepsilon_\perp(\omega)k_z^\mathrm{air}(\omega,\alpha^\mathrm{i})+k_z^\mathrm{SiO{_2},e}(\omega,\alpha^\mathrm{i})},\\
\widetilde{L}_{xx}(2\omega,\alpha^\mathrm{r}) &= \frac{2k_z^\mathrm{air}(2\omega,\alpha^\mathrm{r})}{\varepsilon_\perp(2\omega)k_z^\mathrm{air}(2\omega,\alpha^\mathrm{r}) + k_z^\mathrm{SiO{_2},e}(2\omega,\alpha^\mathrm{r})},\\
\widetilde{L}_{yy}(2\omega,\alpha^\mathrm{r}) &= \frac{2k_z^\mathrm{SiO{_2},o}(2\omega,\alpha^\mathrm{r})}{k_z^\mathrm{air}(2\omega,\alpha^\mathrm{r}) + k_z^\mathrm{SiO{_2},o}(2\omega,\alpha^\mathrm{r})},\\
\widetilde{L}_{zz}(2\omega,\alpha^\mathrm{r}) &= \frac{k_z^\mathrm{SiO{_2},e}(2\omega,\alpha^\mathrm{r})}{\varepsilon_\perp(2\omega)k_z^\mathrm{air}(2\omega,\alpha^\mathrm{r}) + k_z^\mathrm{SiO{_2},e}(2\omega,\alpha^\mathrm{r})}
\label{eq:fresnel}
\end{split}
\end{align}
where $L$ and $\widetilde{L}$ denote the Fresnel factors for the incoming fundamental and outgoing SHG beams, respectively. Here, $\varepsilon_{xx}=\varepsilon_{yy}=\varepsilon_\perp$ and $\varepsilon_{zz}=\varepsilon_\parallel$ are the elements of the diagonal dielectric tensor perpendicular and parallel to the optic ($c$-)axis, respectively, which have been calculated using a multi-oscillator model.\cite{Gervais1975} Also, $k_z^\mathrm{air}(\omega,\alpha^\mathrm{i(r)})=2\pi\omega\cos\alpha^\mathrm{i(r)}$ is the $z$-component of the complex wave vector of the incoming (reflected) waves in air, while $k_z^\mathrm{SiO{_2},o}(\omega,\alpha^\mathrm{i(r)})=2\pi\omega\sqrt{\varepsilon_\perp-\frac{\varepsilon_\perp}{\varepsilon_\parallel}\sin^2\alpha^\mathrm{i(r)}}$ and $k_z^\mathrm{SiO{_2},e}(\omega,\alpha^\mathrm{i(r)})=2\pi\omega\sqrt{\varepsilon_\perp\sin^2\alpha_\mathrm{i(r)}}$ are the $z$-components of the complex wave vectors of the ordinary and extraordinary incoming (reflected) waves inside the quartz sample, respectively.\cite{Mosteller1968} Note the occurrence of the anisotropy factor $\zeta=\varepsilon_\perp/\varepsilon_\parallel$ in $L_{zz}$.\cite{Paarmann2016a}\par 

The trigonal crystal structure of $\alpha$-quartz with point group 32 (Schoenflies $D_3$) results in the following two unique, non-vanishing components of the second-order susceptibility tensor $\chi^{(2)}$ for SHG:\cite{Shen2003}
\begin{align}
\begin{split}
\chi^{(2)}_{aaa} = -\chi^{(2)}_{abb} &= -\chi^{(2)}_{bba} = -\chi^{(2)}_{bab},\\
\chi^{(2)}_{acb} = -\chi^{(2)}_{bac} &= \chi^{(2)}_{abc} = -\chi^{(2)}_{bca}.
\end{split}
\label{eq:chi2_elements}
\end{align}
In contrast, four unique $\chi^{(2)}$ components contribute to SFG\cite{Liu2008} where for instance $\chi^{(2)}_{acb} \neq \chi^{(2)}_{abc}$ and $\chi^{(2)}_{cab} \neq 0$ for two-color visible/IR excitation. Therefore, the higher symmetry of SHG with fewer unique $\chi^{(2)}$ tensor elements will generally make it easier to disentangle the different contributions as compared to SFG.


In order to theoretically describe the azimuthal behavior of the SHG, a transformation of the $\chi^{(2)}$ tensor from the crystal frame into the laboratory frame is required. In general, the $\chi^{(2)}$ tensor elements in the laboratory frame $(x,y,z)$ can be derived from the contributing $\chi^{(2)}$ tensor elements given in terms of crystal coordinates $(a,b,c)$ using\cite{Shen2003}
\begin{equation}
\chi_{ijk}^{(2)}=\sum_{lmn}{\chi_{lmn}^{(2)}(\hat{i}\cdot\hat{l})(\hat{j}\cdot\hat{m})(\hat{k}\cdot\hat{n})},
\label{eq:chi2_transformation}
\end{equation}
where $(\hat{i},\hat{j},\hat{k})$ and $(\hat{l},\hat{m},\hat{n})$ are the basis vectors of the laboratory and crystal frame, respectively. For a $c$-cut crystal and a rotation about the surface normal, the coordinate transformation takes the simple form $\hat{a}=\hat{x}\cos\varphi+\hat{y}\sin\varphi$, $\hat{b}=-\hat{x}\sin\varphi+\hat{y}\cos\varphi$ and $\hat{c}=\hat{z}$, i.e. rotation about $\hat{z}$.\par
Applying the crystal to laboratory frame transformation (Eq.~\ref{eq:chi2_transformation}) and summing over all contributing $\chi^{(2)}$ elements for each given polarization configuration yields the azimuthal behavior of the SHG signal which is non-zero for all possible polarization conditions. Exemplarily, we here show the expressions for the SHG intensity for an SPP polarization geometry (denoting an S-polarized SHG beam and two P-polarized incoming beams) and PPP, respectively:
\begin{widetext}
\begin{align}
\begin{split}
I_\mathrm{SPP}(2\omega,\varphi)\propto &\Big|\widetilde{L}_{yy}(2\omega,\alpha_\text{SHG}^\text{r}) \Big[\overbrace{\big(L_{zz}(\omega,\alpha_1^\mathrm{i})L_{xx}(\omega,\alpha_2^\mathrm{i})+L_{xx}(\omega,\alpha_1^\mathrm{i})L_{zz}(\omega,\alpha_2^\mathrm{i})\big)}^{\equiv L^\mathrm{eff}_{acb}}\chi^{(2)}_{acb}(2\omega,\omega,\omega)\cdots\\
\label{eq:shg_model_SPP}
& + \underbrace{L_{xx}(\omega,\alpha_1^\mathrm{i})L_{xx}(\omega,\alpha_2^\mathrm{i})}_{\equiv L^\mathrm{eff}_{aaa}}\chi^{(2)}_{aaa}(2\omega,\omega,\omega)\sin(3\varphi)\Big]\Big|^2/\Delta k^2,
\end{split}\\
\label{eq:shg_model_PPP}
I_\mathrm{PPP}(2\omega,\varphi)\propto &\Big|\widetilde{L}_{xx}(2\omega,\alpha_\text{SHG}^\text{r})\Big[L_{xx}(\omega,\alpha_1^\mathrm{i})L_{xx}(\omega,\alpha_2^\mathrm{i})\Big]\chi^{(2)}_{aaa}(2\omega,\omega,\omega)\cos(3\varphi)\Big|^2/\Delta k^2.
\end{align}
\end{widetext}
Here, the $3\phi$ dependence is a result of the trigonal symmetry of $\chi^{(2)}_{lmn}$ in Eq.~\ref{eq:chi2_transformation} for a $c$-cut crystal. For example, $\chi^{(2)}_{xxx}=\chi^{(2)}_{aaa}(\cos^3\varphi-3\sin^2\varphi\cos\varphi)=\chi^{(2)}_{aaa}\cos(3\varphi)$. From Eqs.~\ref{eq:shg_model_SPP} and \ref{eq:shg_model_PPP} it becomes clear that for PPP, a 6-fold azimuthal angle-dependence is expected due to the $\chi^{(2)}_{aaa}\cos(3\varphi)$ term, whereas in the SPP configuration the $\chi^{(2)}_{aaa}\sin(3\varphi)$ term interferes with the angle-independent $\chi^{(2)}_{acb}$ term, resulting in a 3-fold azimuthal angle-dependence with relative lobe amplitudes depending on the ratio $\chi^{(2)}_{aaa}/\chi^{(2)}_{acb}$. SHG intensities for other polarization conditions can be derived analogously.

\subsection{Origin of the $\chi^{(2)}$ enhancement}
The $\chi^{(2)}(\omega)$ dispersion in the vicinity of optical phonon resonances has been theoretically described by Flytzanis for zincblende-type crystals which have a single optical phonon resonance.\cite{Flytzanis1972} In his work, he shows that three resonant amplitudes, $C_{1,2,3}$, contribute to the second-order susceptibility, namely the Faust-Henry coefficient\cite{Faust1966} as well as electrical and mechanical anharmonicity,\cite{Flytzanis1972,Roman2006} respectively. In order to apply this model to crystals with multiple phonon resonances like $\alpha$-quartz, a generalized $\chi^{(2)}(\omega)$ expression is required. In analogy to Flytzanis, the multi-oscillator expressions of the two unique tensor elements $\chi^{(2)}_{acb}$ and $\chi^{(2)}_{aaa}$ take the following general form:
\begin{widetext}
\begin{align}
\begin{split}
\chi^{(2)}_{acb}(2\omega,\omega,\omega) = \chi^{(2)}_{\infty ,acb}\bigg[ 1 &+\sum_{j,k} \bigg(\frac{C_1^k}{D_k(2\omega)}+\frac{C_1^j}{D_j(\omega)}+\frac{C_1^k}{D_k(\omega)}\bigg) \cdots \\
& + \sum_j \sum_k \bigg(\frac{C_2^{k,j}}{D_k(2\omega)D_j(\omega)} + \frac{C_2^{k,k}}{D_k(2\omega)D_k(\omega)}+\frac{C_2^{j,k}}{D_j(\omega)D_k(\omega)}\bigg) \cdots\\
& + \sum_k \sum_j \sum_{k^\prime} \frac{C_3^{k,j,k^\prime}}{D_k(2\omega)D_j(\omega)D_{k^\prime}(\omega)} \bigg], \\
\chi^{(2)}_{aaa}(2\omega,\omega,\omega) = \chi^{(2)}_{\infty, aaa}\bigg[ 1 &+\sum_{k} C_1^k \bigg(\frac{1}{D_k(2\omega)}+\frac{1}{D_k(\omega)}+\frac{1}{D_k(\omega)}\bigg) \cdots \\
& + \sum_k \sum_{k^\prime} C_2^{k,k^\prime}\bigg(\frac{1}{D_k(2\omega)D_{k^\prime}(\omega)} + \frac{1}{D_k(2\omega)D_{k^\prime}(\omega)}+\frac{1}{D_{k^\prime}(\omega)D_{k^\prime}(\omega)}\bigg) \cdots\\
& + \sum_k \sum_{k^\prime} \sum_{k^{\prime\prime}} \frac{C_3^{k,k^\prime,k^{\prime\prime}}}{D_k(2\omega)D_{k^\prime}(\omega)D_{k^{\prime\prime}}(\omega)} \bigg],
\end{split}
\label{eq:chi2_model}
\end{align}
\end{widetext}
where $D_i(\omega)=1-\omega^2/\left(\Omega^\mathrm{TO}_i\right)^2-\mathrm{i}\gamma_i\omega/\left(\Omega^\mathrm{TO}_i\right)^2$ is the resonant denominator of the $i$th phonon resonance and indices $j$ and $k^{(^\prime,^{\prime\prime})}$ run through the extraordinary and ordinary phonon modes of quartz, respectively. 

According to Flytzanis and later work by Roman et~al.~\cite{Roman2006}, the resonant amplitudes can be written in our generalized model as:
\begin{align}
C_1^k = \frac{\alpha^k_\mathrm{TO}}{2 v \chi^{(2)}_\infty}\left( \frac{Z^*}{M \omega_\mathrm{TO}^k}\right),
\end{align}
which is the Faust-Henry coefficient of mode $k$,
\begin{align}
C_2^{k,j} = \frac{\mu_{k,j}^{(2)}}{2 v \chi^{(2)}_\infty}\left( \frac{Z^*}{M \omega_\mathrm{TO}^k}\right)^2,
\end{align}
and 
\begin{align}
C_3^{k,j,k'} = \frac{\phi_{k,j,k'}^{(3)}}{2 v \chi^{(2)}_\infty}\left( \frac{Z^*}{M \omega_\mathrm{TO}^k}\right)^3.
\end{align}
Here, $Z^*$ is the effective charge, $v$ the volume of the unit cell, and $M$ the reduced mass. The three important parameters in these equations are the polarizability $\alpha_\mathrm{TO}^{k}$, the electrical anharmonicity $\mu_{k,j}^{(2)}$, and the mechanical anharmonicity $\phi_{k,j,k'}^{(3)}$. Notably, SFG spectroscopy only probes the first-order polarizability, i.e. Raman term,\cite{Liu2008} because of the different selection rules for mixed visible-IR excitation which is singly resonant in the IR response, i.e. does not probe the anharmonicity of vibrational potentials. In contrast, IR-only SHG is doubly resonant in the IR response and thereby provides access to mechanical and electrical anharmonicities of vibrational modes and does not require Raman-type interaction. In particular the latter was argued to be dominant for III-V semiconductors\cite{Roman2006} and was experimentally shown to be significant for the single-mode polar crystal SiC.\cite{Paarmann2016a} Specifically for anisotropic multimode materials like $\alpha$-quartz, these higher order anharmonic terms are of particular importance since they contain information about anharmonic coupling between the different phonon modes. For instance for $\chi^{(2)}_{acb}$, Eq.~\ref{eq:chi2_model} predicts only cross terms between planar ($k$) and axial ($j$) phonons to contribute to incoming resonances in the higher order resonance terms. Therefore, the quantification of these resonant amplitudes would provide a unique experimental access to anharmonic mode coupling in polar crystals.


\section{Experimental Results}

\subsection{\label{sec:room_temp}SHG phonon spectrum at room temperature}
\begin{figure*}[htb]
\includegraphics[width=0.95\textwidth]{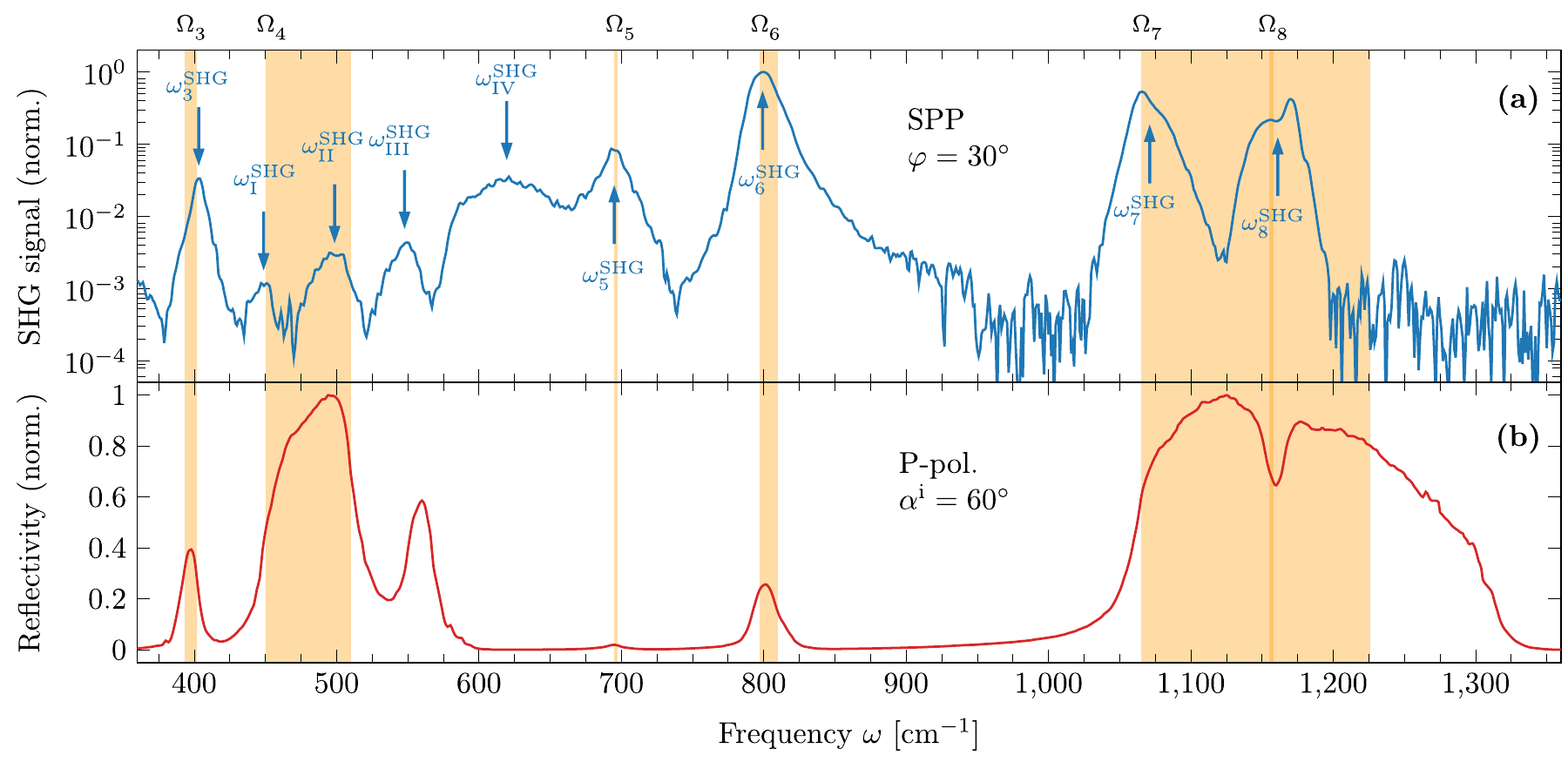}
\caption{(a): Experimental SHG spectrum of $\alpha$-quartz for SPP polarization conditions at an azimuthal angle $\varphi=\ang{30}$. Strong SHG enhancements at TO phonon frequencies are observed. (b): Reflectivity spectrum for P-polarization measured at an angle of incidence $\alpha^\mathrm{i}=\ang{60}$ with respect to the surface normal, demonstrating the correlation between nonlinear and linear spectroscopy. Shaded orange boxes mark the phonon modes, ranging from the respective lower frequency TO phonons to the corresponding higher frequency LO phonons, indicating the formation of Reststrahlen regions. SHG peaks appear at spectral positions marked by blue arrows and are labeled $\omega^\mathrm{SHG}_j$.}
\label{fig:shg_room_temp}
\end{figure*}

In Fig.~\ref{fig:shg_room_temp}a, we show the experimental SHG spectrum of $\alpha$-quartz for the SPP polarization geometry measured at room temperature at an azimuthal angle $\varphi=\ang{30}$. 
We observe sharp resonances ranging over about three orders of magnitude at spectral positions $\omega^\mathrm{SHG}_j$ that can be attributed to transversal optical (TO) phonons of $\alpha$-quartz, most prominently at frequencies $\Omega^\mathrm{TO}_j$ of the IR-active $E$-type TO phonon modes as indicated by the left borders of the orange shades in Fig.~\ref{fig:shg_room_temp} (values taken from Ref.~\onlinecite{Gervais1975}). The strong enhancements are primarily due to a combination of resonances in the nonlinear susceptibility $\chi^{(2)}$ (Eq.~\ref{eq:chi2_model}) as well as in the local field magnitudes which enter in the form of Fresnel transmission tensor elements (Eq.~\ref{eq:fresnel}) and the wavevector mismatch $\Delta k$ in Eq.~\ref{eq:sh_intensity}. The interplay of these highly dispersive quantities makes up the essence of mid-IR SHG as a phonon spectroscopy. The spectral features in the range from \SIrange{445}{620}{\per\centi\meter}, labeled I-IV, cannot be unambiguously assigned to $E$-type phonon resonances, due to the very low signal levels and unusual temperature dependence, see  Sec.~\ref{sec:temp_data}. Unlike the strong TO phonon resonances, longitudinal optical (LO) phonons only cause subtle signatures in quartz's SHG spectrum as discussed in detail in Sec.~\ref{sec:shg_origin}, where we analyze the SHG spectrum quantitatively.\par
Simultaneous reflectivity measurements at an angle of incidence $\alpha^\mathrm{i}=\ang{60}$ reveal two distinct regions of particularly high reflectance between corresponding TO and LO phonon frequencies, i.e. $\Omega^\text{TO(LO)}_4$ and $\Omega^\text{TO(LO)}_7$, respectively (see orange shade in Fig.~\ref{fig:shg_room_temp}). These so-called Reststrahlen bands emerge for strong modes between the respective TO and LO phonon frequencies where the real part of the dielectric function takes negative values, resulting in strongly attenuated evanescent waves and thus high reflectivity.\cite{Adachi1999} The other phonon modes have a smaller TO-LO frequency splitting and thus smaller oscillator strengths,\cite{Adachi1999} such that no full Reststrahlen bands are formed. Instead, less intense peak-like features appear in the reflectivity spectrum.\par

\begin{figure*}[htb]
\includegraphics[width=0.96\textwidth]{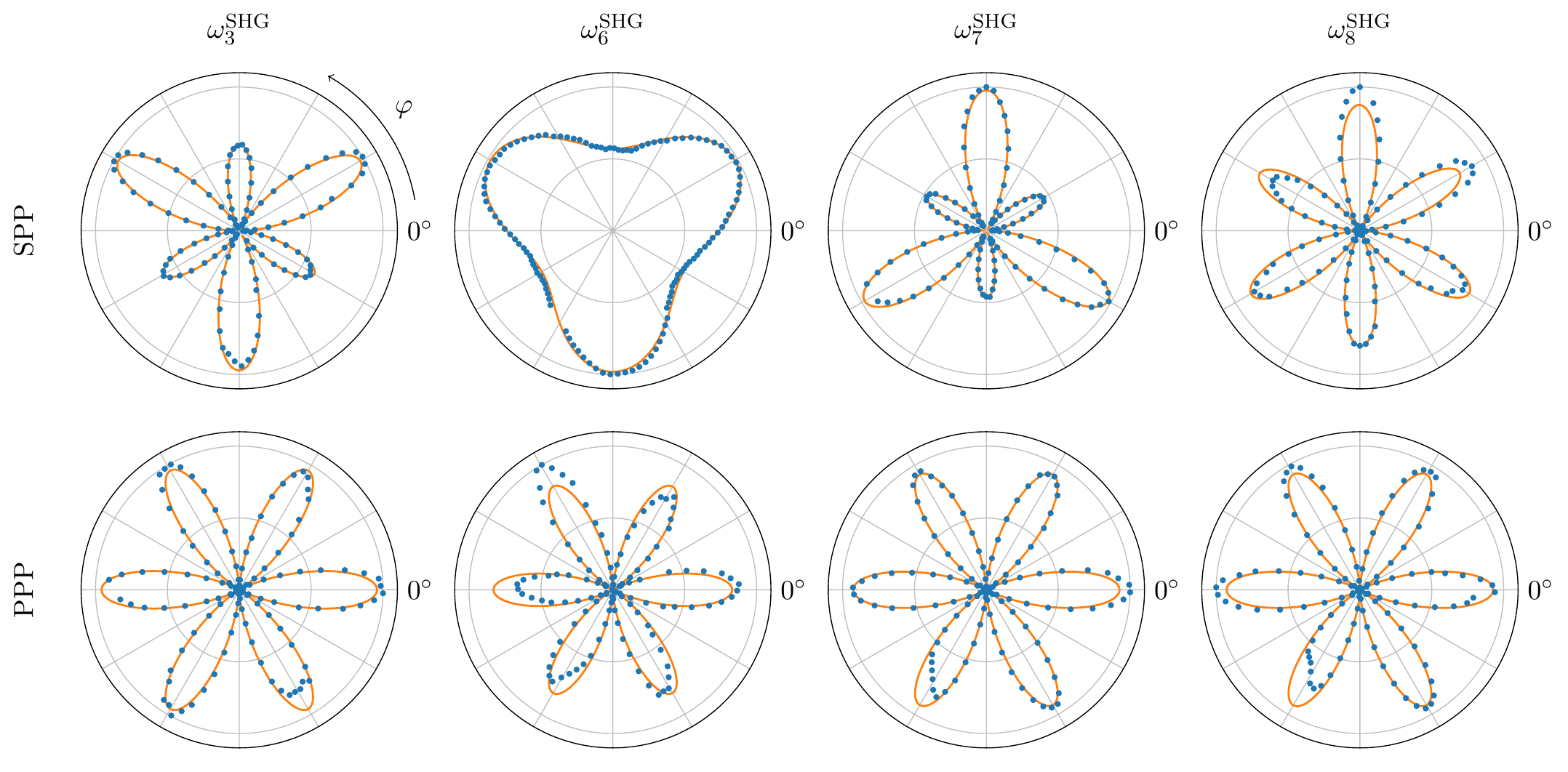}
\caption{Azimuthal behavior of the SHG at four exemplary TO phonon resonances for SPP and PPP polarization conditions. Solid lines represent model fits using Eqs.~\ref{eq:shg_model_SPP} and \ref{eq:shg_model_PPP}, respectively. While PPP measurements exhibit a clear sixfold symmetry, SPP measurements show a threefold symmetry due to the two uniquely contributing $\chi^{(2)}$ elements, $\chi^{(2)}_{acb}$ and $\chi^{(2)}_{aaa}$.}
\label{fig:azimuth_room_temp}
\end{figure*}

Additionally, we have measured the azimuthal behavior of the SHG intensity at room temperature at spectral positions $\omega^\mathrm{SHG}_j$ of all SHG resonances that are marked in Fig.~\ref{fig:shg_room_temp} for two different polarization conditions, SPP and PPP, at room temperature. Fig.~\ref{fig:azimuth_room_temp} shows the azimuthal scans at four exemplary spectral positions, i.e. at $\omega^\text{SHG}_3$, $\omega^\text{SHG}_6$, $\omega^\text{SHG}_7$ and $\omega^\text{SHG}_8$. Model fits (solid lines) using Eqs.~\ref{eq:shg_model_SPP} and \ref{eq:shg_model_PPP} are in good agreement with the experimental data. While SPP scans depend on both uniquely contributing $\chi^{(2)}$ tensor elements, $\chi^{(2)}_{aaa}$ and $\chi^{(2)}_{acb}$, PPP scans solely depend on $\chi^{(2)}_{aaa}$, resulting in the observed threefold and sixfold symmetric azimuthal behavior as expected from theory, Eqs.~\ref{eq:shg_model_SPP} and \ref{eq:shg_model_PPP}, respectively. The quantitative knowledge of the Fresnel factors as described in section \ref{sec:theory}, allows for extraction of relative magnitudes of the tensor elements $\chi^{(2)}_{acb}/\chi^{(2)}_{aaa}$ from SPP scans (Table~\ref{tab:chi2_ratio}).


\begin{table}[htb]
\caption{Fit parameters from all acquired SPP azimuthal scans. Quantitative knowledge of the Fresnel transmission coefficients allows for an extraction of the ratio $\chi^{(2)}_{acb}/\chi^{(2)}_{aaa}$.}
\begin{ruledtabular}
\begin{tabular}{cdD{,}{\pm}{6.5}}
\textrm{Label $\omega^\text{SHG}_j$} & \multicolumn{1}{c}{\textrm{$\omega$} [\si{\per\centi\meter}]} & \multicolumn{1}{c}{$\chi^{(2)}_{acb}/\chi^{(2)}_{aaa}$} \\
\hline \\ [-2.0ex]
3 & 400  &  0.186,0.014\\
5 & 690  &  0.22,0.05\\
6 & 795  &  1.46 ,0.19\\
7 & 1069 &  2.6 ,0.3\\
8 & 1171 &  0.11 ,0.08 \\
\hline \\ [-2.0ex]
I   & 445 &  0.32,0.05\\
II  & 498 &  30  ,4\\
III & 545 &  2.5    ,1.2\\
IV	& 620 &  0.654,0.014
\end{tabular}
\end{ruledtabular}
\label{tab:chi2_ratio}
\end{table}

It should be noted that the azimuthal behavior of the TO6 mode in PPP polarization is highly sensitive to slight misalignment of the detection polarizer, which leads to an interference of the S-polarized SHG component with the P-polarized signal. Because of the relatively large isotropic component ($\propto L^\mathrm{eff}_{acb}\chi^{(2)}_{acb}$, cf. Eq. \ref{eq:shg_model_SPP}) entering the SPP signal for this phonon and a generally stronger signal for SPP than for PPP, a small polarization angle offset in detection has a significant impact on the measured azimuthal behavior. This effect becomes critical especially at the TO6 mode where the SPP signal is non-zero at angles where the PPP component also yields signal. For the other measured phonon resonances, however, this interference is less pronounced since SPP signals are zero at angles where PPP yields the largest SHG signal.

\subsection{\label{sec:temp_data}Temperature-dependence of the SHG phonon spectrum}

In order to investigate the temperature sensitivity of second-harmonic phonon spectroscopy, measurements using a sample heating stage have been conducted. In these measurements, the $\alpha\rightarrow\beta$ phase transition of quartz was of particular interest. During this displacive phase transition at nominally $T_\text{c}=\SI{846}{\kelvin}$,\cite{Gervais1975} quartz changes from the trigonal crystal symmetry to a hexagonal one (point group 622, Schoenflies $D_6$). In the $\beta$-phase, the $E$-type phonon modes labeled $\Omega_3$, $\Omega_5$ and $\Omega_8$ become IR-forbidden due to structural changes in the crystal.\cite{Scott1967}\par
\begin{figure*}[htb]
\includegraphics[width=0.95\textwidth]{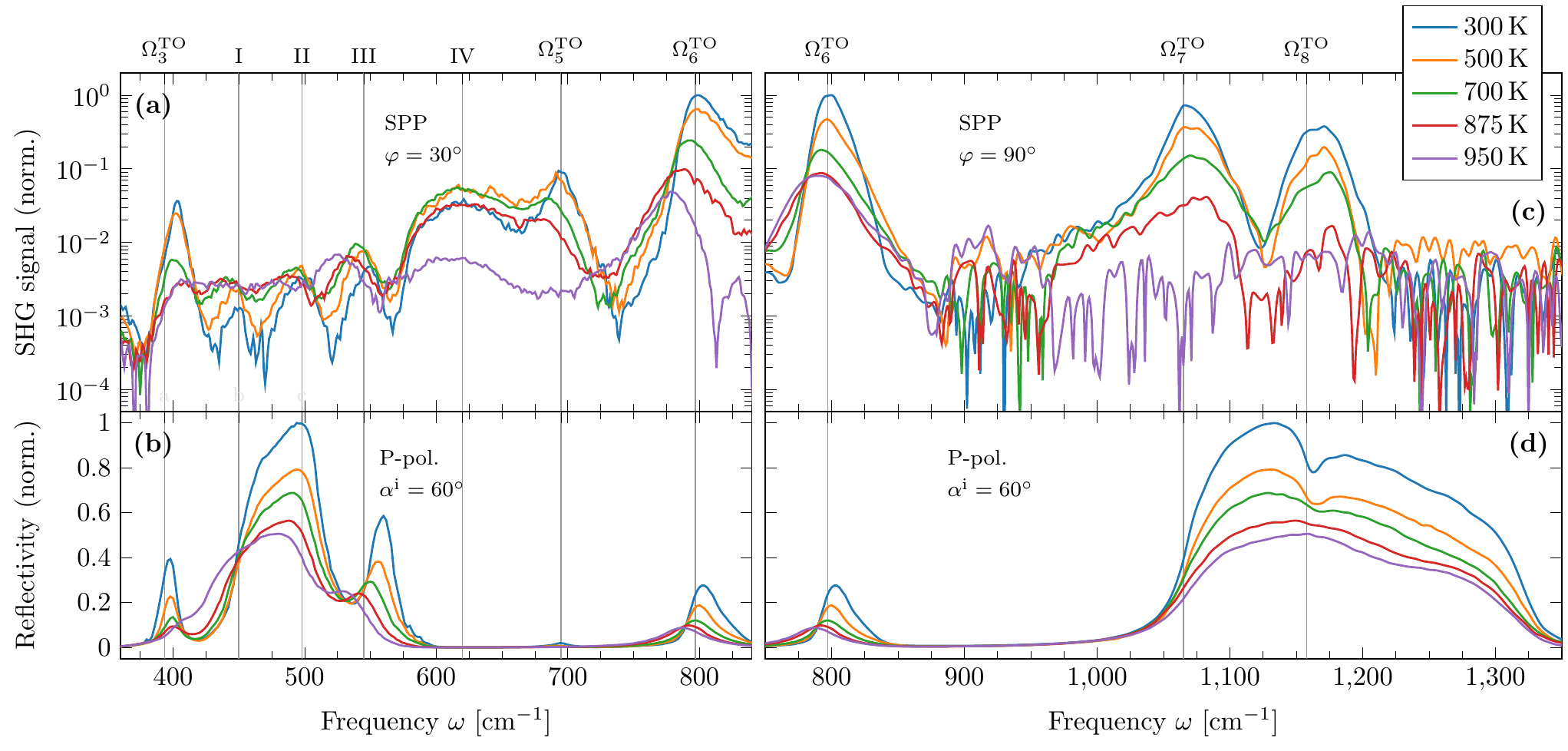}
\caption{Temperature-dependent SHG and reflectivity spectra for temperatures ranging from room temperature up to \SI{950}{\kelvin}. Due to the different azimuthal shapes of the SHG in SPP polarization, these data were taken at $\varphi=\ang{30}$ in the low-frequency range (a,b) and at $\varphi=\ang{90}$ in the high-frequency range (c,d). The SHG resonances in (a,c) decrease and broaden, while spectral positions shift with higher temperatures. Note the logarithmic scale.  The corresponding reflectivity spectra (b,d) were taken at an angle of incidence of \ang{60} and in P-polarization. Spectral features in the reflectivity behave in accordance with their respective SHG peaks.}
\label{fig:shg_temp}
\end{figure*}
Fig. \ref{fig:shg_temp}a shows SHG spectra measured in SPP polarization at \SI{300}{\kelvin}, \SI{500}{\kelvin}, \SI{700}{\kelvin}, \SI{875}{\kelvin} and \SI{950}{\kelvin}. Notably, the different azimuthal behavior of the modes, see Fig.~\ref{fig:azimuth_room_temp},  limited the dynamic range for some modes, e.g. TO3 or TO7, if the spectra were taken at a single azimuth. Therefore, we acquired the data in two parts: at azimuthal angle $\varphi=30^\circ$ in the low-frequency region from \SIrange{350}{850}{\per\centi\meter}, and at $\varphi=90^\circ$ in the high-frequency region from \SIrange{750}{1350}{\per\centi\meter}, see Fig.~\ref{fig:shg_temp} (a,b) and (c,d), respectively. 

From the temperature-dependent spectra, two observations become apparent. Firstly, a clear decrease of the resonant amplitudes (SHG peak height) as well as increased dampings (SHG peak width) at higher temperatures are observed. Secondly, SHG peak positions shift with temperature. This behavior is due to temperature-dependent changes of the phonon damping rates, $\gamma^\mathrm{TO}$, and frequencies, $\Omega^\text{TO}$. 
Remarkably, some TO phonon resonances show a particularly sudden drop in SHG intensity close to the nominal phase transition at $T_\text{c}=\SI{846}{\kelvin}$, most notably TO5, while other resonant features exhibit a gradual decrease up to the highest measured temperature, e.g. TO6, or until the SHG signal falls below detection level, e.g. TO7. This behavior likely originates from the $\alpha\rightarrow\beta$ phase transition, where the phonon modes TO3, TO5 and TO8 become IR-inactive, while TO6 and TO7 remain active.\cite{Gervais1975}\par

The corresponding reflectivity spectra are shown in Fig.~\ref{fig:shg_temp} (b,d). Like the peaks in the SHG spectra, Reststrahlen edges and peaks in the reflectivity soften and shift spectrally as temperature increases. Near the phase transition temperature, the peak-like features associated with the TO3 and TO5 modes nearly disappear as does the dip-like feature in the upper Reststrahlen band at the TO8 mode frequency, thereby being consistent with the SHG data. Note that the reflectivity data are plotted on a linear scale as opposed to the SHG spectra which are plotted logarithmically to cover the large dynamic range in these signals. This representation of the data masks the fact that the SHG peak heights are much more susceptible to phonon damping than any of the reflectivity features.\par

Notably, in the frequency range from $\sim$\SIrange{420}{570}{\per\centi\meter}, the SHG intensity of modes I-IV even increases with higher sample temperature. The plateau-like feature labeled IV shows some peculiar behavior with a pronounced signal drop above the phase transition temperature. While these observations are very interesting and not understood at this stage, we also note that we observed a signal contribution due to black-body radiation by FEL-induced heating of the sample, getting more pronounced at elevated sample temperatures. This effect has been accounted for by taking background spectra that were measured at a large time delay between the pulses in both excitation arms and subtracted from the SHG signals. Still, due to the low signal levels in this frequency range, the possibility of a thermal contribution to the measured SHG signal cannot be ruled out entirely.



\section{Analysis and Discussion}
A quantitative analysis of the SHG spectra is challenging due to the numerous resonances that introduce a multitude of fit parameters to the $\chi^{(2)}$ description discussed in Sec.~\ref{sec:theory} (Eq.~\ref{eq:chi2_model}). Furthermore, strongly temperature-dependent damping rates cause a rapid drop of SHG intensities at high temperatures which makes it difficult to assess features close to and above $T_\mathrm{c}$. Nevertheless, we here attempt to fit the SHG spectrum at room temperature by simplifying the fit model described above to a reduced number of parameters. For the temperature-dependent spectra, we empirically fit the SHG peaks in order to perform a quantitative analysis of the observed features, particularly with regard to quartz's $\alpha\rightarrow\beta$ phase transition.

\subsection{\label{sec:shg_origin}Origin of the SHG enhancement}

\begin{figure*}[htb]
\includegraphics[width=1\textwidth]{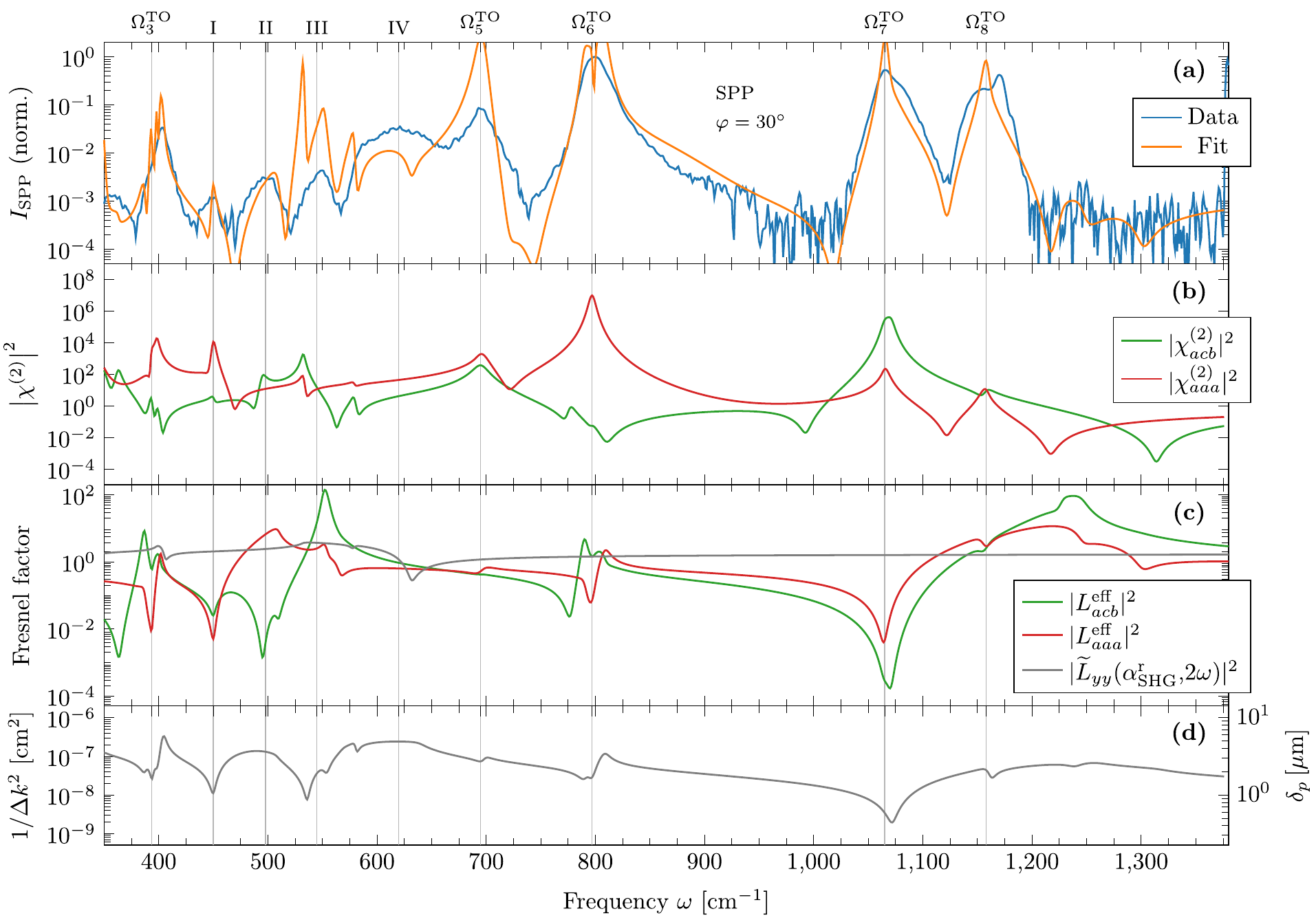}
\caption{Composition of the measured SHG spectrum: The highly dispersive $\chi^{(2)}$ tensor elements (b), the Fresnel factors (c) as well as the inverse squared wave vector mismatch $1/\Delta k^2$ (d) enter the measured SHG signal (a). The latter also determines an effective escape depth $\delta_p$ of the SHG light (d). Note the logarithmic scales in all four graphs.}
\label{fig:shg_dispersion}
\end{figure*}

The origin of the observed SHG enhancement is threefold. First, the highly dispersive second-order nonlinear susceptibility $\chi^{(2)}(\omega)$ typically peaks at TO phonon frequencies, causing a strong enhancement of the SHG signal over several orders of magnitude.\cite{Paarmann2015} Secondly, the signals are strongly modulated by the dispersing Fresnel factors. The latter can be accurately calculated using Eq.~\ref{eq:fresnel}, and we plot the results in Fig.~\ref{fig:shg_dispersion}c for both contributing terms in SPP geometry. Finally, also the wave vector mismatch $\Delta k$ is strongly dispersive as shown in Fig.~\ref{fig:shg_dispersion}d, defining the effective escape depth $\delta_p=1/\Delta k$ of the SHG light. For non-absorbing materials in a reflective geometry, this is typically about half the wavelength of the SHG light\cite{Shen2003} and therefore largely non-dispersive. Here, however, the dispersion of $\Delta k^2$ arises due to strong absorption of the TO phonons, i.e. through the large imaginary part of $k_{1,2}^\mathrm{SiO_2}$ at phonon resonances dominating $\Delta k^2$ in these cases. The resulting effective escape depth, $\delta_p$, is also shown in Fig.~\ref{fig:shg_dispersion}d.

For the dispersion of $\chi^{(2)}(\omega)$ with the numerous phonon resonances in quartz, the generalization of Flytzanis's model as described above (Eq.~\ref{eq:chi2_model}) results in a very large number of independent parameters $C_{1,2,3}^{j,k}$. Therefore, in order to apply a model fit to the measured data, we simplify the generalized model by reducing the fit parameters to $C_1$ terms as well as $C_3$ cross-terms which only contain spectrally proximate phonon resonances, as those are assumed to be the main contributing terms to the $\chi^{(2)}(\omega)$ function. This leaves a reduced number of 19 independent $C_{1,3}$ coefficients which enter as fit parameters. This strongly simplified model, applied to Eq.~\ref{eq:shg_model_SPP}, while lacking quantitative accuracy, already reproduces nearly every feature observed in the SHG spectrum at least qualitatively as shown by the fit in Fig.~\ref{fig:shg_dispersion}a (orange line). From the fit results, it is possible to extract the substantial spectral shapes of the contributing $\chi^{(2)}(\omega)$ functions (plotted in Fig.~\ref{fig:shg_dispersion}b).\par

Fig.~\ref{fig:shg_dispersion} illustrates the interplay of all dispersing contributions to the SHG signals, i.e. the $\chi^{(2)}(\omega)$ (Fig.~\ref{fig:shg_dispersion}b), Fresnel dispersion (Fig.~\ref{fig:shg_dispersion}c) as well as the wave vector mismatch (Fig.~\ref{fig:shg_dispersion}d), resulting in the SHG spectrum (Fig.~\ref{fig:shg_dispersion}a). Here, compensating effects become apparent as can be clearly seen by taking the TO7 mode as an example: Fresnel factors and the wave vector mismatch suppress the SHG signal significantly ($\sim 10^{-3}$ and $\sim 10^{-1}$, respectively) such that the $\chi^{(2)}$ enhancement ($\sim 10^{6}$) must outdo this effect in order to allow for a measurable SHG peak at $\Omega_7^\mathrm{TO}$. 
Furthermore, Fresnel factors can cause a spectral shift of the SHG peak position relative to the phonon frequency as observed at the TO3 mode where a Fresnel suppression of the SHG signal at $\Omega_3^\mathrm{TO}$ and a simultaneous enhancement at a slightly higher frequency ($\sim\SI{403}{\per\centi\meter}$) cause this offset between SHG peak position, $\omega^\mathrm{SHG}_3$, and phonon frequency, $\Omega_3^\mathrm{TO}$, showing that a thorough treatment of the linear optical effects is necessary when interpreting the SHG spectra.\par
However, and in contrast to previous work on single-mode SiC\cite{Paarmann2016a}, we here only observe subtle signatures of LO phonon resonances in the SHG data, owing to a combination of effects. First, weaker modes, e.g. $\Omega_5$ or $\Omega_8$, reside on the tails of strong mode resonances in the dielectric function, strongly suppressing Fresnel resonances at their LO frequencies. Secondly, out-of-plane Fresnel resonances ($L_{zz}$) are generally found to be stronger than in-plane resonances ($L_{xx}$ and $L_{yy}$), see Fig.~\ref{fig:shg_dispersion}c. Since for trigonal $\alpha$-quartz, we are mostly sensitive to $L_{xx}$ and $L_{yy}$ (see Eqs.~\ref{eq:shg_model_SPP} and \ref{eq:shg_model_PPP}), Fresnel resonance effects are reduced as compared to, e.g., hexagonal SiC.\cite{Paarmann2016a} However, peaks at $\omega^\mathrm{SHG}_3$, $\omega^\mathrm{SHG}_\mathrm{II}$ and $\omega^\mathrm{SHG}_6$ are likely to be, at least in part, due to LO phonon resonances.

\begin{figure*}[htb]
\includegraphics[width=0.9\linewidth]{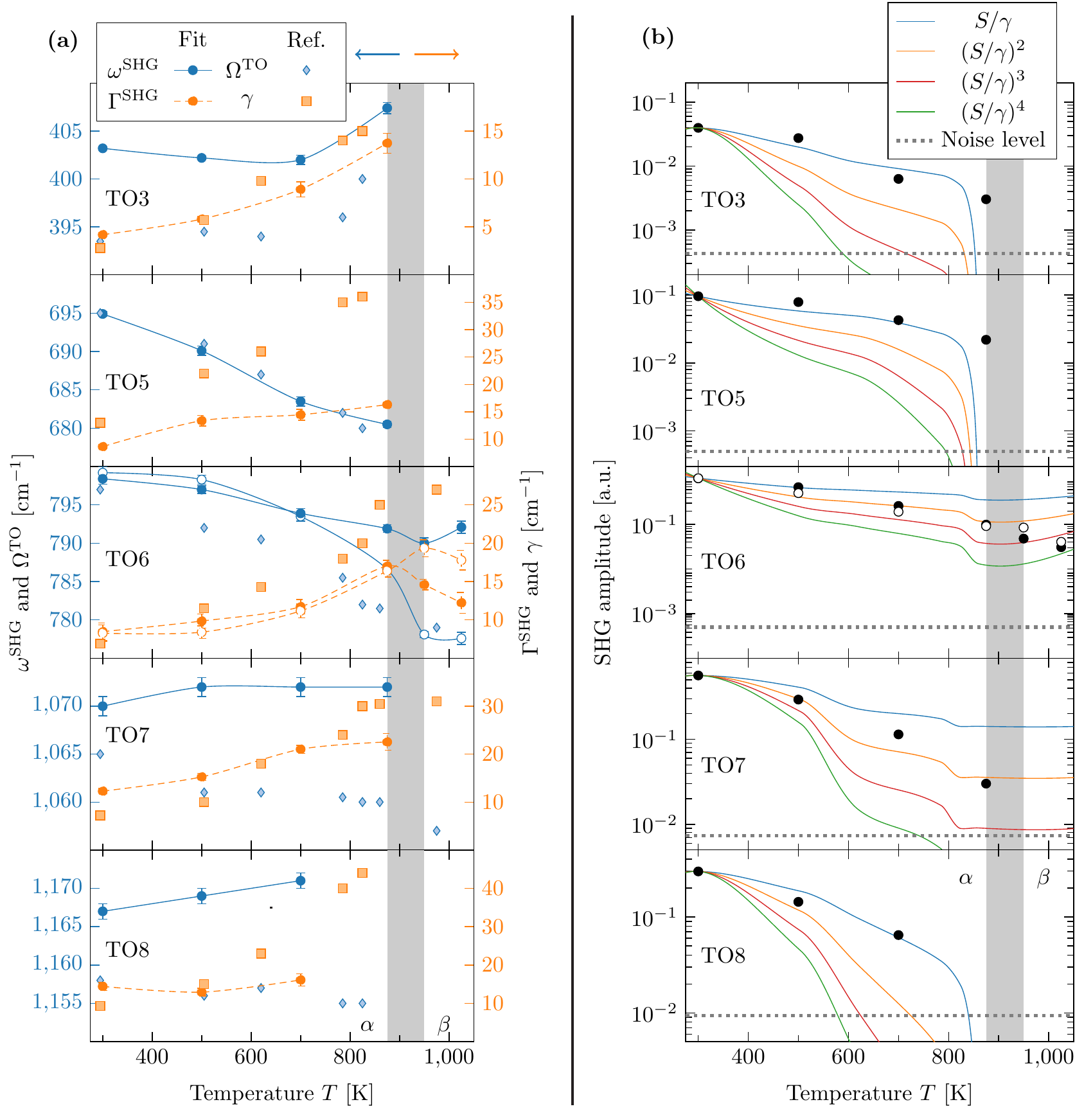}
\caption{(a): Fitted peak positions (blue) and damping constants (orange) from the SHG spectrum (dots with error bars) compared to TO phonon frequencies and dampings, respectively, as measured by Gervais and Piriou\cite{Gervais1975} (diamonds and squares). Lines are a guide to the eye. The $\alpha\rightarrow\beta$ phase transition of quartz presumably takes place in the temperature interval indicated by the gray shaded area. (b): Fitted SHG peak amplitudes at phonon resonances (black dots) with estimations of their temperature-dependent behavior of the SHG based on the respective oscillator stengths $S$ and damping constants $\gamma$. For TO6 in (a) and (b), the closed and open dots correspond to the data taken at $\varphi=\ang{30}$ and $\ang{90}$ in Fig.~\ref{fig:shg_temp}a and c, respectively.}
\label{fig:comparison}
\end{figure*}

\subsection{\label{sec:temp_analysis}Temperature dependence of the phonon resonances}

For a quantitative analysis of the observed spectral features, especially at high temperatures, we compare the SHG peak positions and widths with temperature-dependent phonon data acquired by linear IR reflectivity measurements by Gervais and Piriou.\cite{Gervais1975} Thereto, resonant features in the SHG spectra have been fitted with a Lorentzian function, yielding the center frequencies and line widths, $\omega^\mathrm{SHG}$ and $\Gamma^\mathrm{SHG}$ (half width at half maximum, HWHM), respectively, as well as their amplitudes. The fit results for all temperatures are plotted in Fig.~\ref{fig:comparison}a together with the phonon data as measured in Ref. \onlinecite{Gervais1975} for comparison.

In Fig.~\ref{fig:comparison}a it can be seen how the TO phonon frequencies, $\Omega^\mathrm{TO}$, and damping constants, $\gamma$, compare to the peak characteristics observed in the SHG spectrum. The SHG peak positions mimic the temperature-dependent trends of their respective phonon frequencies reasonably well, although offsets of up to \SI{10}{\per\centi\meter} are observed due to the influence of Fresnel factors and $\Delta k^2$ as discussed above. This is also apparent in the data for TO6 which was taken for two values of the azimuthal angle, $\varphi=\ang{30}$ and $\ang{90}$ (Fig.~\ref{fig:comparison}), shown as closed and open dots, respectively, since here the two relevant Fresnel components $L^\mathrm{eff}_{acb}$ and $L^\mathrm{eff}_{aaa}$ in Eq.~\ref{eq:shg_model_SPP}, both having a different temperature dependence, contribute differently to the total SHG signal.

The effect of phonon damping constants directly translates to SHG peak widths, in large parts even in remarkable quantitative agreement with $\Gamma^\mathrm{SHG}$. This is non-trivial, as the SHG spectra are modulated by the highly dispersive Fresnel factors, especially at phonon resonances. Still, the data in Fig.~\ref{fig:comparison}a shows that the SHG peak widths provide a reasonably accurate estimation of phonon damping constants for all modes.\par


Fig.~\ref{fig:comparison}b, on the other hand, shows the fitted SHG peak amplitudes. 
The data clearly shows a continuous decrease of SHG intensity with higher temperatures for all resonances, with all but the TO6 resonance amplitudes vanishing above the phase transition temperature.  The amplitude of the latter exhibits---as its corresponding spectral position and width---a kink-like behavior at the phase transition. 

In order to estimate the expected behavior of the resonant SHG amplitudes with temperature, we evaluated the temperature dependence of the amplitude of each phonon resonance to $S/\gamma$, where $S$ is the oscillator strength and $\gamma$ the damping of the respective phonon, using the multi-oscillator model\cite{Adachi1999} with data from Ref.~\onlinecite{Gervais1975}. Since $S/\gamma$ enters the the Fresnel factors (Eq.~\ref{eq:fresnel}) through the resonances in the dielectric function to different orders, and similarly, it also contributes to $\chi^{(2)}(\omega)$ (Eq.~\ref{eq:chi2_model}) linearly and quadratically, we plot several powers $(S/\gamma)^N$, $N=1-4$, alongside the data in Fig.~\ref{fig:comparison}b. Here, we normalized $(S/\gamma)^N$ to match the experimental data at room temperature. 

Below the phase transition, the experimental amplitudes linearly follow $S/\gamma$ for TO3, TO5, and TO8, while the amplitudes of TO6 and TO7 appear to decay more quickly. Near the phase transition temperature $T_\mathrm{c}=846$~K, expected in the range marked by the gray-shaded area in Fig.~\ref{fig:comparison}, $S/\gamma$ rapidly drops to zero for the modes TO3, TO5, and TO8, since these modes become IR-forbidden in the $\beta$-phase such that their oscillator strength $S$ vanishes.  Indeed, no SHG signal is observed for these modes above $T_\mathrm{c}$. For TO6 and TO7, on the other hand, $S/\gamma$ predicts appreciable amplitudes also above $T_\mathrm{c}$ since these modes persist through the phase transition. In the experiment, we do indeed observe TO6 above $T_\mathrm{c}$, while TO7 also vanishes---in contradiction to our expectation. However, a careful examination of noise levels (dotted line in Fig. \ref{fig:comparison}b), reveals that the amplitude of TO7 in fact would fall below detection level as crossing the \SI{875}{\kelvin} measurement point, assuming that it shares the same trend with $S/\gamma$ as TO6, which continues to drop more rapidly than predicted by the power laws above $T_\mathrm{c}$.
It is striking that those modes which happen to persist through the phase transition exhibit a temperature-dependent behavior distinct from the other modes. Although this empirical observation is very interesting, it is at this point not possible to isolate a single cause for this effect due to the numerous contributions to the $\chi^{(2)}$ line shape.

\subsection{Discussion}
A quantitative analysis of the SHG spectra turned out to be a challenging task, in particular if two contributions with many phonon resonances interfere as for the data in Fig.~\ref{fig:shg_room_temp}, leading to a number of fit parameters so large that quantitative fitting is not feasible. With sufficient signal-to-noise, a potential solution to this problem could be achieved by making use of the symmetry properties of Eq.~\ref{eq:shg_model_SPP}. The data in Fig.~\ref{fig:shg_room_temp} were recorded at an azimuthal angle $\varphi=30^\circ$, i.e. at maximum signal magnitude for many of the resonances, cf. Fig.~\ref{fig:azimuth_room_temp}. If instead one would acquire spectra at $\varphi=0^\circ$, one would exclusively probe $\chi^{(2)}_{acb}$, and could divide out the linear quantities $L$ and $\Delta k$ to directly measure the $\chi^{(2)}$ line shape. At room temperature this is challenging due to low signal levels, however, at low temperatures, $< \SI{100}{\kelvin}$, such an experiment becomes feasible. Similarly, SHG spectra taken in PPP geometry would exclusively yield the line shape of $\chi^{(2)}_{aaa}$. Respective experiments are currently under way and will provide direct access to anharmonic mode coupling. 

At high temperatures where the $\alpha\rightarrow\beta$ phase transition takes place, however, such an approach is not applicable. Due to the highly temperature-dependent damping rates, signals around $T_\mathrm{c}$ are generally weak and for some resonances even fall below detection level. In consequence, temperature-dependent azimuthal scans are not sufficiently meaningful.\cite{Winta2016} Otherwise, those would have been the tool of choice to study the structural phase transition: considering $\beta$-quartz's hexagonal crystal symmetry where no azimuthal dependence is expected in contrast to the trigonal $\alpha$-phase (cf. Fig.~\ref{fig:azimuth_room_temp}), we expected extreme sensitivity of our technique to the $\alpha\rightarrow\beta$ phase transition, which unfortunately was completely washed out by the weak SHG signals at high temperatures.

However, when comparing resonant features in the temperature-dependent SHG spectra to their respective $\chi^{(2)}_{acb}/\chi^{(2)}_{aaa}$ ratios at room temperature (Tab.~\ref{tab:chi2_ratio}), one correlation stands out: $\beta$-forbidden phonon resonances consistently exhibit a $\chi^{(2)}_{acb}/\chi^{(2)}_{aaa}$ ratio smaller than 1, while for TO6 and TO7 which persist through the phase transition, $\chi^{(2)}_{acb}/\chi^{(2)}_{aaa}>1$. In fact, the SHG intensity for $\beta$-quartz (point group 622, Schoenflies $D_6$) in SPP geometry reads:
\begin{equation}
I_\mathrm{SPP}^\beta(2\omega)\propto \Big|\widetilde{L}_{yy}(2\omega,\alpha_\text{SHG}^\text{r}) L^\mathrm{eff}_{acb} \chi^{(2)}_{acb}(2\omega,\omega,\omega)\Big|^2/\Delta k^2,
\label{eq:shg_beta}
\end{equation}
which is identical to the second term in Eq.~\ref{eq:shg_model_SPP}, i.e., the SHG component that---according to the azimuthal---dominates the SHG for $\beta$-allowed phonon resonances ($\propto\chi^{(2)}_{acb}$) at room temperature. Note, however, that for $\beta$-forbidden modes, both $\chi^{(2)}$ symmetry components vanish upon entering the higher symmetry of the $\beta$-phase. This suggests that already at room temperature the azimuthal scans give an indication of which phonon modes will persist above $T_\mathrm{c}$ and which will not. It remains to be shown how general this effect is, and if a similar behavior is found for other phase transitions, for instance in ferroelectrics or multiferroics.\cite{Fiebig2005,Fiebig2016}\par
Our observation reflects the structural and symmetry information intrinsically contained in the resonant contributions of the $\chi^{(2)}$ tensor elements, potentially making SHG phonon spectroscopy an excellent spectroscopic technique for studies of phase transitions. This holds true especially for phase transitions that take place below room temperature where damping rates are typically low, therefore allowing generally larger SHG signals and making azimuthal scans a promising tool to study symmetry changes during phase transitions. In fact, the sensitivity to symmetry changes makes an obvious case for SHG over linear spectroscopy. Additionally, the pronounced peaks in the SHG make phonon resonances considerably easier to extract than, e.g., from Reststrahlen edges in IR reflectivity spectra\cite{Paarmann2015} while its nonlinearity causes an improved contrast.

\section{Conclusion}

We have demonstrated second-harmonic phonon spectroscopy as a highly sensitive tool to study phonon resonances in non-centrosymmetric polar crystals, using $\alpha$-quartz as a model system. Its second-order nonlinearity makes it very sensitive to phonon resonances which can be detected across several orders of magnitude. It opens up additional experimental degrees of freedom compared to linear techniques which can be exploited using polarization control to selectively access information related to the crystal symmetry as seen in the azimuthal behavior of the SHG signal via the $\chi^{(2)}(\omega)$ tensor elements. Furthermore, second-harmonic phonon spectroscopy has also been shown to be temperature-sensitive, allowing to track phonon frequencies and linewidths as well as detecting symmetry changes in the sample across a structural phase transition.

\section*{Acknowledgments}
The authors thank R. Kramer Campen for helpful discussions and careful reading of the manuscript.

%

\end{document}